\documentclass[aip, apl,nolongbibliography,reprint]{revtex4-2}
\usepackage{lingmacros}
\usepackage{tree-dvips}
\usepackage{graphics}
\usepackage{graphicx}
\graphicspath{{figs/}}
\usepackage[caption=false]{subfig}
\usepackage{subfig}
\usepackage{epsfig}
\usepackage{epsf,epic}
\usepackage{color}
\usepackage{marvosym }
\usepackage{siunitx}
\usepackage{amsmath}
\usepackage{amssymb}
\usepackage{amsfonts}
\usepackage{wrapfig}
\usepackage{pstricks}
\usepackage{multirow}
\usepackage{pst-node}
\usepackage{extdash}
\usepackage{braket}
\usepackage{bm}
\usepackage{dcolumn}
\newcommand{\etal}{\textit{et al.\ }}

\newcommand{\mc}{\multicolumn}

\begin{document}

\title{Band offsets at  the interfaces between $\beta$-Ga$_2$O$_3$ and Al$_2$O$_3$}
\author{Sai Lyu}
\email{sailyu@sdnu.edu.cn}
\affiliation{School of Physics and Electronics, Shandong Normal University, Jinan 250358, China}
\affiliation{Shandong Provincial Engineering and Technical Center of Light Manipulations, Shandong Normal University, Jinan, 250358, China}

\begin{abstract}
The band offsets and the chemical bonding at the interfaces between ($\bar{2}$01) $\beta$-Ga$_2$O$_3$ and Al$_2$O$_3$ polymorphs are studied through hybrid functional calculations. For alumina, we consider four representative phases, i.e., $\alpha$, $\kappa$, $\theta$ and $\gamma$-Al$_2$O$_3$. We generate realistic slab models for the interfaces which satisfy electron counting rules. The O atoms bridge the $\beta$-Ga$_2$O$_3$ and the Al$_2$O$_3$ slabs and all the dangling bonds are saturated. The band offsets are obtained by applying an alignment scheme which requires separate bulk and interface calculations. The calculated band offsets are useful for the design of devices based on the $\beta$-Ga$_2$O$_3$/Al$_2$O$_3$ heterojunctions, particularly $\beta$-Ga$_2$O$_3$ metal-oxide semiconductor field effect transistors. 

\end{abstract}
\maketitle


Gallium oxide (Ga$_2$O$_3$) is a promising candidate to advance existing technologies in the field of high-power electronics and solar-blind ultraviolet (UV) photodetectors because of its large band gap. \cite{Pearton2018} Among the five identified polymorphs of Ga$_2$O$_3$, $\beta$-Ga$_2$O$_3$ has the most stable crystal structure and thus has attracted a great deal of recent attention. \cite{Pearton2018}
This material has a wide band gap of 4.5-4.9 eV and its high breakdown electric field significantly exceeds that of the commonly used SiC and GaN. \cite{Higashiwaki2012,Mastro2017}  Most importantly, bulk crystals of $\beta$-Ga$_2$O$_3$ can be produced from the melt by using melt growth techniques at a potentially lower cost than the fabrications of SiC and GaN.\cite{Pearton2018, Kuramata2016}


In the development of electronic devices based on $\beta$-Ga$_2$O$_3$, the fabrication of metal-oxide semiconductor field effect transistors (MOSFETs) has  been recently demonstrated. \cite{Chabak2016,Higashiwaki2013,Hwang2014,Ahn2016} For $\beta$-Ga$_2$O$_3$ MOSFETs, a semiconductor with a high dielectric constant (high $\kappa$) is desirable to serve as a gate dielectric so as to reduce the device operating voltage.\cite{Kamimura2014,Carey2017} Moreover, a gate dielectric must provide sufficient barriers for both electrons and holes, which requires a sufficiently large band gap to obtain the desired band offsets($\gtrsim$ 1 eV). \cite{Carey2017} Al$_2$O$_3$ has been identified as a good candidate because of its large band gap and high dielectric constant. \cite{Pearton2018,Gitzen2006, Dorre2011} 
Recently, Kamimura \etal obtained a conduction band offset (CBO) of 1.5 eV and a corresponding valence band offset (VBO) of 0.7 eV at the $\alpha$-Al$_2$O$_3$/$\beta$-Ga$_2$O$_3$ (010) interface.\cite{Kamimura2014} In Ref.\ \onlinecite{Carey2017}, the VBO was measured to be 0.07 eV for atomic layer deposited (ALD) $\alpha$-Al$_2$O$_3$ on ($\bar{2}$01) $\beta$-Ga$_2$O$_3$ and $-$0.86 eV for sputtered $\alpha$-Al$_2$O$_3$ on Ga$_2$O$_3$. And the corresponding CBO was measured to be 2.23 eV and 3.16 eV, respectively.  Hung \etal found a CBO of 1.7 eV on atomic layer deposited Al$_2$O$_3$/Ga$_2$O$_3$ ($\bar{2}$01) interface through capacitance-voltage measurements. \cite{Hung2014} Hattori \etal measured VBO of 0.5 eV and the CBO of 1.9 eV, respectively, at the $\gamma$-Al$_2$O$_3$/$\beta$-Ga$_2$O$_3$ (010) interface. \cite{Hattori2016} 

Band offsets are critical parameters for designs of heterostructures. However, the reported values for both VBO and CBO at the Al$_2$O$_3$/$\beta$-Ga$_2$O$_3$ interfaces clearly exhibit a large variability. Take the VBO at the Al$_2$O$_3$/$\beta$-Ga$_2$O$_3$ interface as an example, the reported value vary as much as 1.5 eV. Such an ambiguity is also found
for some other dielectrics deposited on $\beta$-Ga$_2$O$_3$.\cite{Pearton2018} 
Some possible reasons include interface disorder, surface termination, and so on.\cite{Pearton2018}
Given this, there is a clearly a need to elucidate the atomic structures and the chemical bondings at these interfaces.
Besides, most of the recent studies are limited to the $\alpha$ phase, without considering other phases which also have large band gaps and high dielectric constants. \cite{Gitzen2006, Dorre2011}
 Therefore, computational investigations are necessary to accurately determine the band offsets between $\beta$-Ga$_2$O$_3$ and Al$_2$O$_3$ polymorphs.

In this work, we study the interfaces between $\beta$-Ga$_2$O$_3$ and Al$_2$O$_3$ using density functional theory (DFT). We investigate four representative phases of Al$_2$O$_3$, i.e., $\alpha$, $\theta$, $\kappa$, and $\gamma$. To overcome the band gap problem, we use hybrid density functional to determine the electronic band structure. The band offsets are obtained through an alignment scheme in which bulk calculations and interface calculations are combined. \cite{Alkauskas2008}
Our study can provide guide for future synthesis and device design, especially for the design of $\beta$-Ga$_2$O$_3$ MOSFETs.


Our DFT calculations employ the Gaussian plane waves (GPW) method as implemented in the CP2K code. \cite{cp2k}The GPW method can efficiently solve the Kohn-Sham equation\cite{KohnSham} by using Gaussians as basis set and plane waves (PW) as auxiliary basis. We use double-$\zeta$ basis sets\cite{VandeVondele2007} and Goedecker–Teter–Hutter (GTH) \cite{Goedecker1996}  pseudopotentials for all the atoms. Treating the Ga $3d$ electrons as valence is important to appropriately describe its electronic band structure. The energy cutoff of PW expansion is 600 Ry and the Brillouin zone is sampled by the $\Gamma$ point when a sufficiently large supercell is used in the calculations. For direct-gap semiconductors, their gaps are calculated using sufficient large supercells. For indirect-gap semiconductors, their gaps are determined with a k-point mesh. The geometry optimizations use the generalized gradient approximation in the form of Perdew-Burke-Ernzerhof  (PBE).\cite{pbegga} The established experimental band gaps of $\beta$-Ga$_2$O$_3$ and Al$_2$O$_3$ are reproduced through an common approach of adjusting the fractions $\alpha$ of Fock exchange in the PBE0($\alpha$) hybrid functionals. \cite{Perdew1996, Adamo1999} In the PBE0($\alpha$) calculations, auxiliary density matrix method adopted to accelerate the time-consuming Fock exchange calculations.\cite{Guidon2010}

The band offsets at the interfaces are determined through the alignment procedure described in Ref.\ \onlinecite{VandeWalle1986,Baldereschi1988, Alkauskas2008}. 
For a heterojunction $A/B$, this procedure requires an interface calculation and two separate bulk calculations for bulk components $A$ and $B$. To be more specific, 
the VBO of a heterojunction $A/B$ is calculated from the following equation: 
\begin{eqnarray}
\textrm{VBO} (A/B)=&(&E_{\textrm {VBM}}^{B}-\bar{V}^{B} )-(E_{\textrm {VBM}}^{A}-\bar{V}^{A} )+  \nonumber  \\
 &(&\bar{V}^{B}-\bar{V}^{A}) 
\end{eqnarray}
where $E_{\textrm{VBM}}-\bar{V}$ is the valence band maximum (VBM) with respect to the bulk reference level determined in two separate bulk calculations, and $\bar{V}^{B}-\bar{V}^{A}$ is the interface lineup of bulk reference levels determined in the interface calculation. We follow the common practice of choosing the averaged electrostatic potential as the  bulk reference level. 
The corresponding CBO can then be calculated from the following equation:  
\begin{equation}
\textrm{CBO} (A/B)=(E_{\textrm g}^{B} -E_{\textrm g}^{A} )+\textrm{VBO}(A/B)
\end{equation}
where $E_{\textrm g}$ denotes the band gap of each interface component. 
The interface lineup is obtained at the GGA level which can yield almost the same interface lineup as hybrid functionals but be less computationally expensive. \cite{Weston2018,Shaltaf2008, Alkauskas2008, Steiner2014} To obtain the interface lineup, we first calculate the $xy$ planar average of the electrostatic potential and then apply a double convolution along the $z$ direction which is  perpendicular to the interface plane. \cite{Baroni1989book,Baldereschi1988} In the interface models, the asymmetric slabs give rise to finite electric fields across the interfaces under the periodic boundary conditions.
\cite{Bengtsson1999-hv}  
To eliminate the effects of the electric fields on the interface lineup, we adopt the extrapolation scheme developed by Foster \etal \cite{Foster2014} In this scheme, the macroscopically averaged electrostatic potential for each interface component is extrapolated from its bulklike region to the nominal interface position. Herein we take the midway between the surface Ga layer and Al layer as the nominal interface position. The interface lineup is obtained by calculating the difference between two extrapolations at the nominal interface position. This extrapolation scheme has been successfully applied to the $\beta$-Ga$_2$O$_3$/AlN and $\beta$-Ga$_2$O$_3$/GaN interfaces. \cite{Lyu2020}

\begin{figure}[ht]
\hspace*{-0.10cm}
\subfloat[$\alpha$-Al$_2$O$_3$ \label{}]{
\includegraphics[width=4cm]{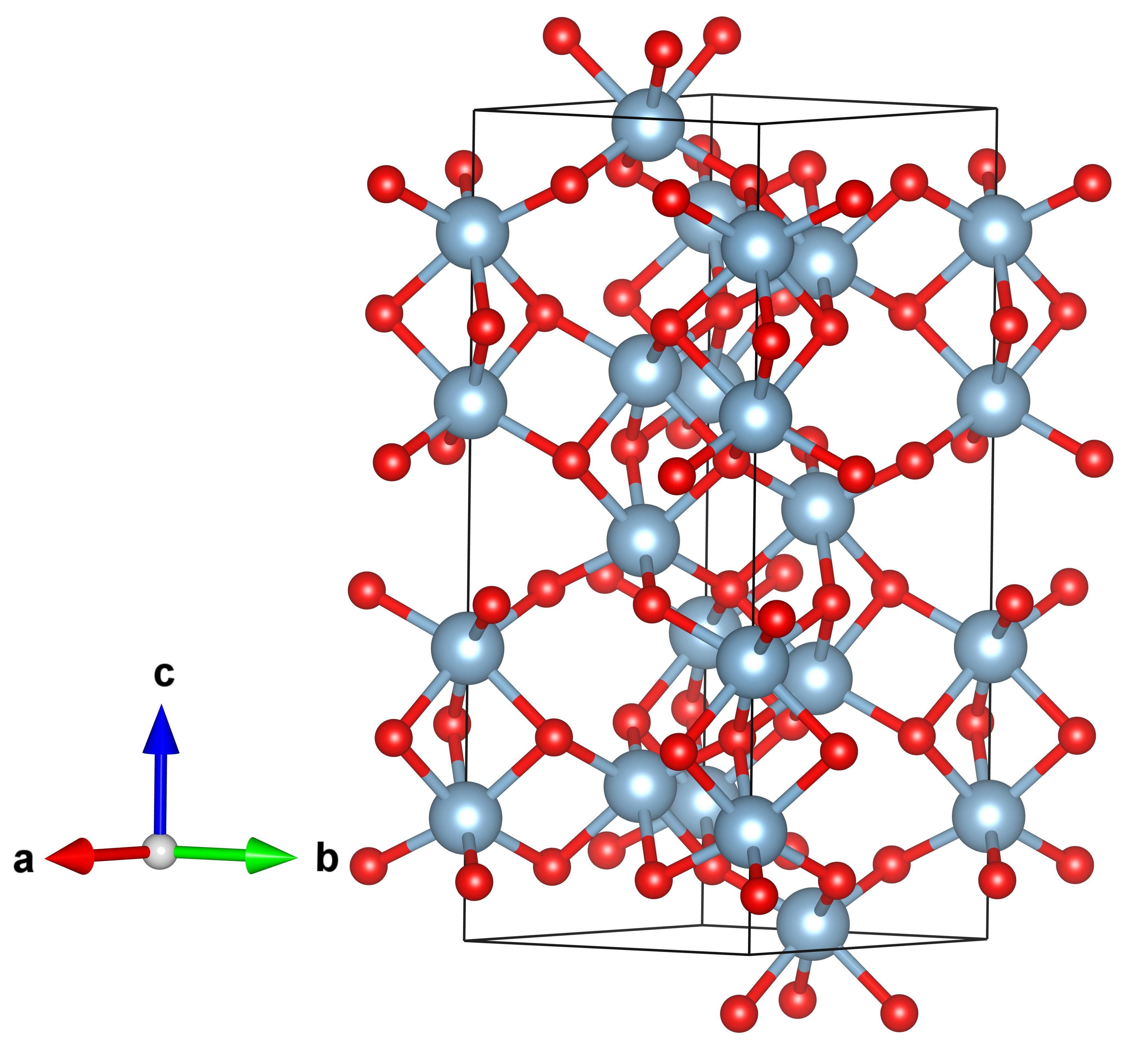}
}
\subfloat[$\kappa$-Al$_2$O$_3$ \label{}]{
\includegraphics[width=4cm]{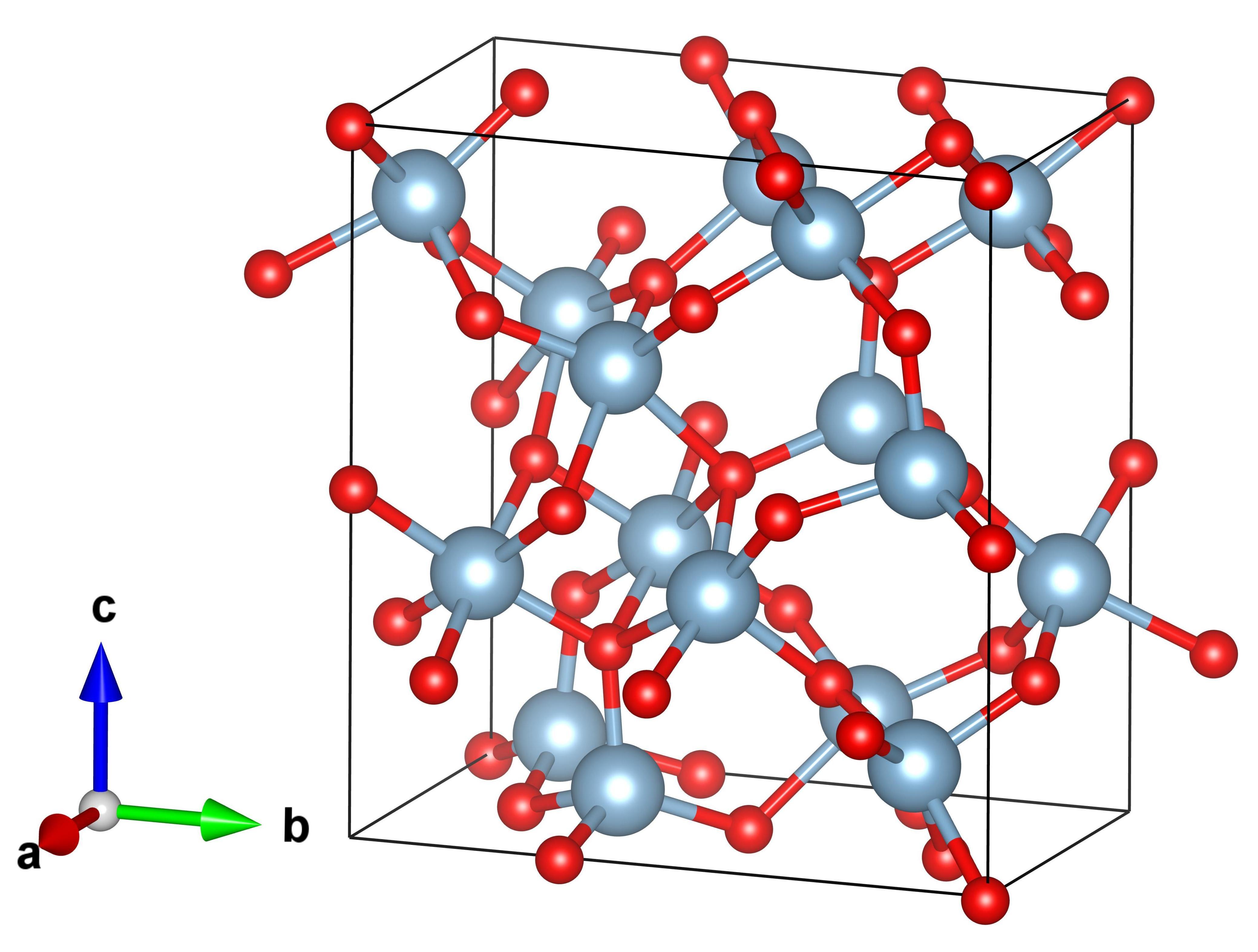} 
} \\
\subfloat[$\theta$-Al$_2$O$_3$ \label{}]{
\includegraphics[width=3.6cm]{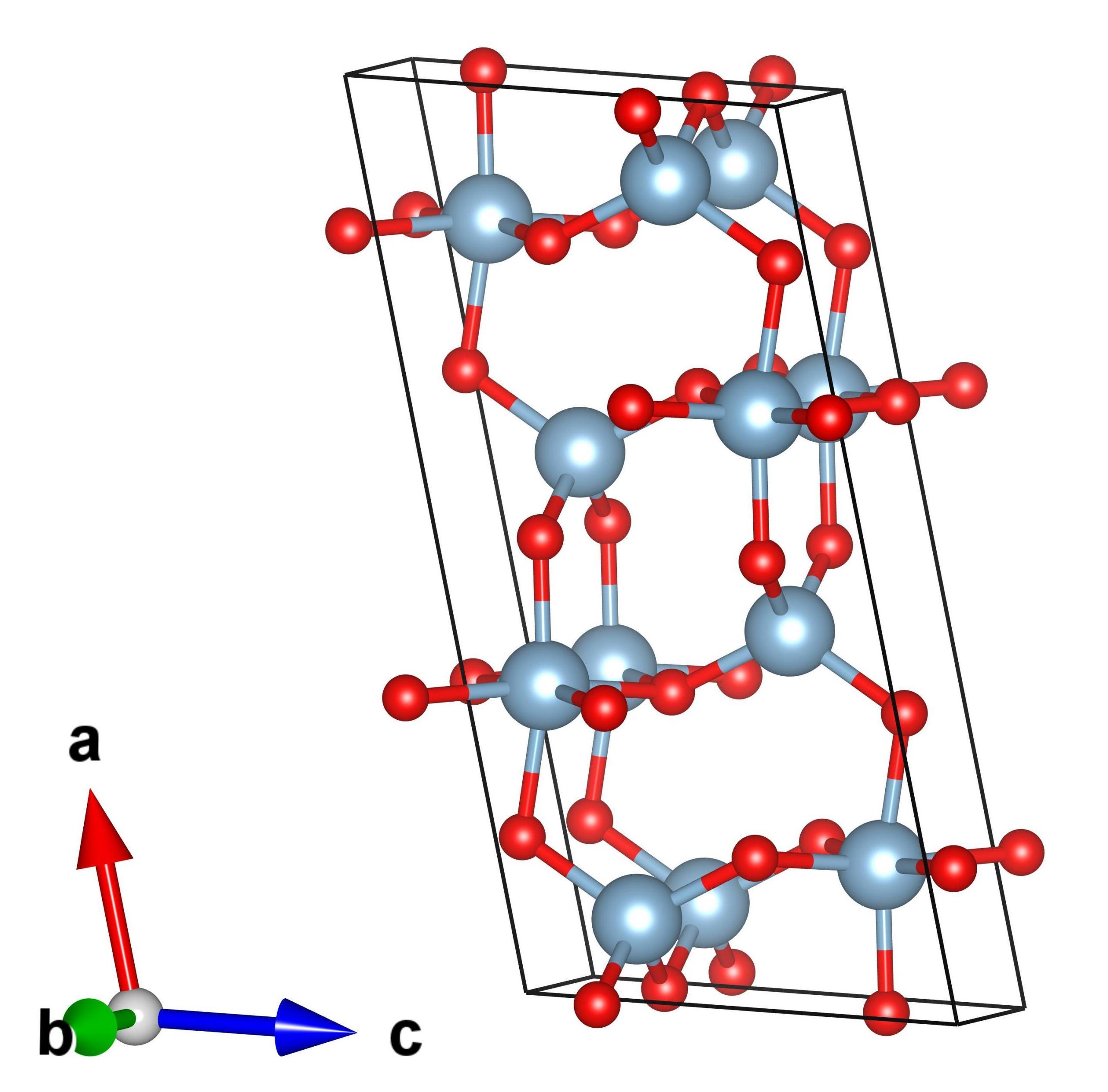} 
}
\subfloat[$\gamma$-Al$_2$O$_3$ \label{}]{
\includegraphics[width=4cm]{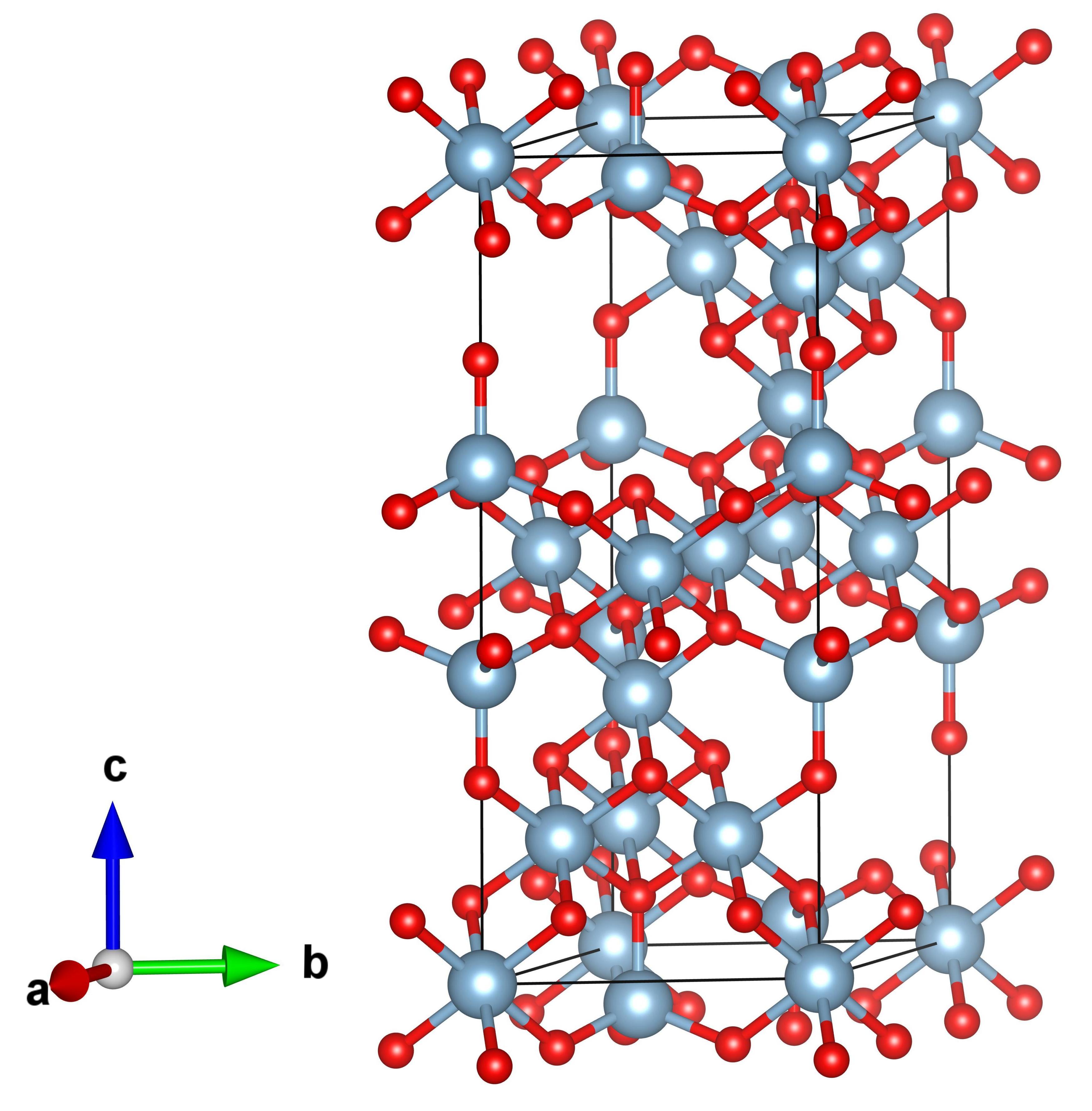} 
} \\
\subfloat[$\beta$-Ga$_2$O$_3$ \label{}]{
\includegraphics[width=4cm]{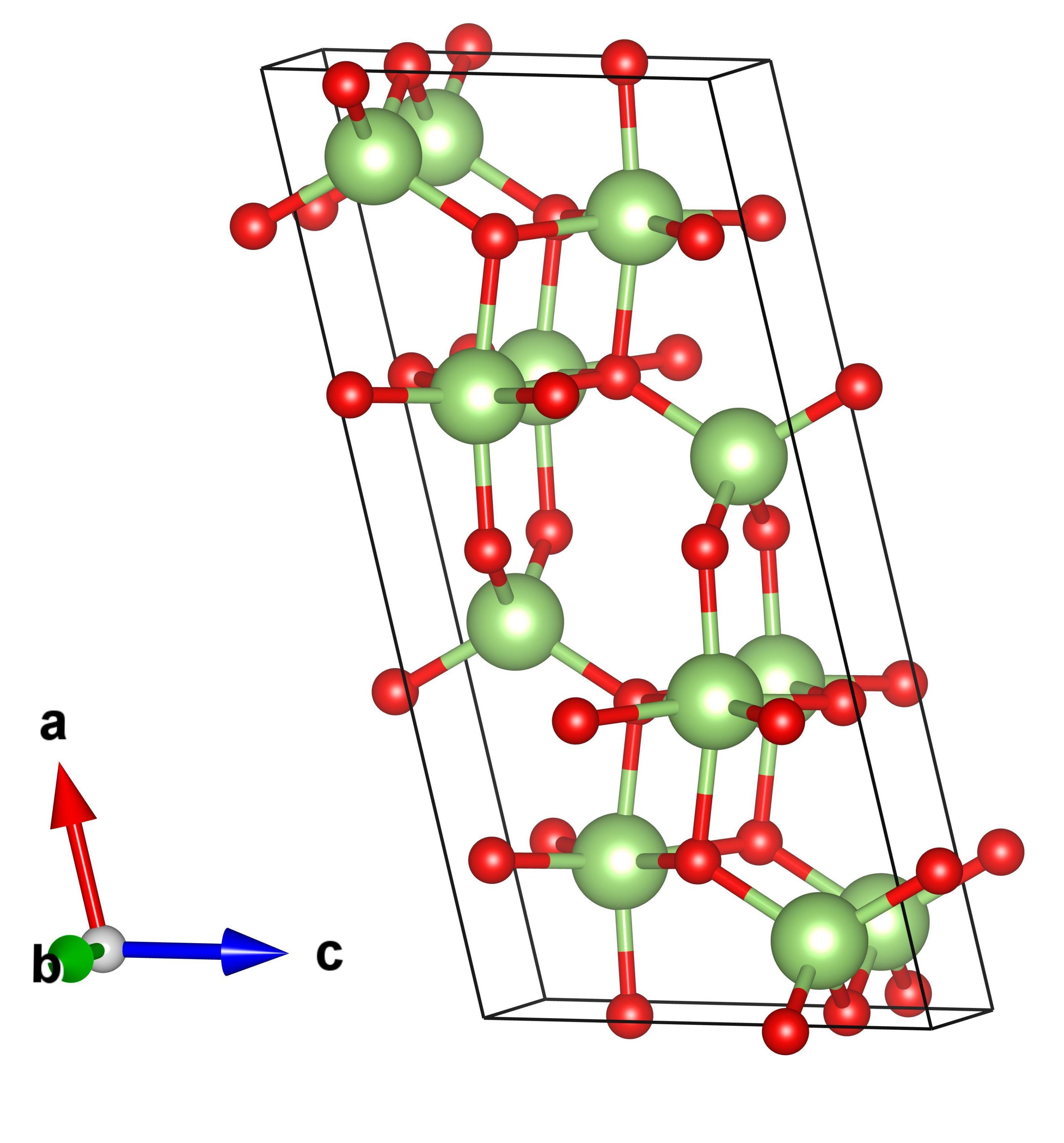} 
}
\caption{Unit cells of $\beta$-Ga$_2$O$_3$ and Al$_2$O$_3$. (a) $\alpha$-Al$_2$O$_3$ (hexagonal), (b) $\kappa$-Al$_2$O$_3$ (orthorhombic), (c) $\theta$-Al$_2$O$_3$ (monoclinic), (d) $\gamma$-Al$_2$O$_3$ (hexagonal), and (e) $\beta$-Ga$_2$O$_3$ (monoclinic). The red, green, and grey spheres indicate O, Ga, and Al atoms, respectively.  \label{unitcell}}
\end{figure}
In Fig.\ \ref{unitcell}, we show the units cells of  the four phases of Al$_2$O$_3$ ($\alpha$, $\kappa$, $\theta$, and $\gamma$) and $\beta$-Ga$_2$O$_3$ studied in this work. 
For a structural model of $\alpha$-Al$_2$O$_3$,  the Al cations occupy the octahedral sites and the O anions are in the vertices of octahedrons. Its space group belongs to $R\bar{3}c$. When represented by a hexagonal lattice as shown in Fig.\ \ref{unitcell} (a), $\alpha$-Al$_2$O$_3$ contains alternative Al and O layers. 
In the case of $\kappa$-Al$_2$O$_3$, the Al cations occupy either octahedral sites or tetrahedra sites surrounded by the O anions. The crystal structure of $\kappa$-Al$_2$O$_3$ corresponds to the space group $Pna2_1$ in the orthorhombic class.\cite{Ollivier1997, Yourdshahyan2004}.
Monoclinic $\theta$-Al$_2$O$_3$ has a space group of $C2/m$ with the Al cations on either octahedral and tetrahedra sites.\cite{Zhou1991}
The model of $\theta$-Al$_2$O$_3$ is based on the crystal structure determined in Ref.\ \onlinecite{Zhou1991}.  
For $\gamma$-Al$_2$O$_2$, we use a 40-atom hexagonal cell comprising eight Al$_2$O$_3$ units. The O anions sublattice is fully occupied and two Al octahedral sites are unoccupied which are farthest from each other.  
This model is derived from the cubic spinel structure with a lattice constant of 7.94 $\textrm\AA$ refined in Ref.\ \onlinecite{Smrcok2006}, and for more details of the model construction, we refer to Refs.\ \onlinecite{Yazdanmehr2012,Wolverton2000}. The experimental lattice constants $a$ and $c$ for this hexagonal model are derived to be 5.61 $\textrm\AA$ and 13.75 $\textrm\AA$, respectively. 
$\beta$-Ga$_2$O$_3$ has a monoclinic crystal structure with the Ga cations belonging to either distorted tetrahedra or distorted octahedra.\cite{Geller1960, Mastro2017} It has the same space group with $\theta$-Al$_2$O$_3$, i.e., $C2/m$, making it easily form alloys with $\theta$-Al$_2$O$_3$.\cite{Peelaers2018, Amol2020} The lattice parameters of the bulk $\beta$-Ga$_2$O$_3$ and the four phases of Al$_2$O$_3$ are obtained through fully geometry optimizations with the GGA functional, which are summarized in Table\ \ref{tablatt}. The corresponding experimental lattice parameters and band gaps are also given.
The band gaps and VBM positions with respect to the bulk reference levels are obtained through PBE0($\alpha$) calculations. 

\begin{figure}[ht]
\hspace*{-0.10cm}
\subfloat[$\alpha$-Al$_2$O$_3$/$\beta$-Ga$_2$O$_3$ interface]{
\includegraphics[width=8cm]{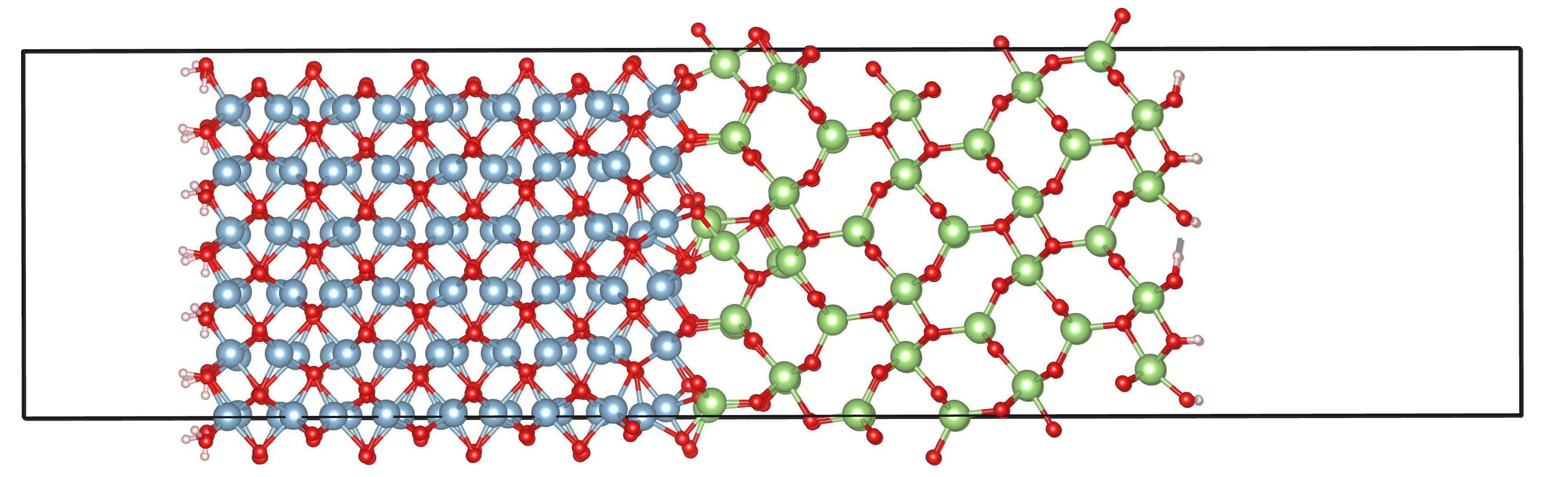} 
}
\\
\subfloat[$\kappa$-Al$_2$O$_3$/$\beta$-Ga$_2$O$_3$ interface]{ 
\includegraphics[width=8cm]{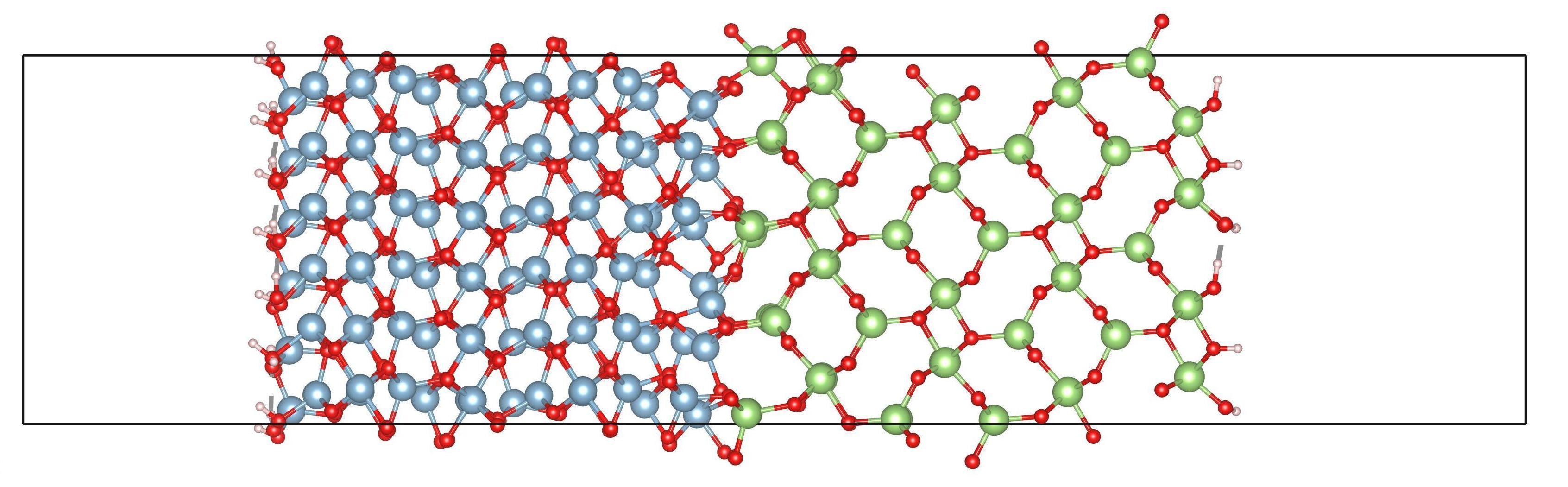} 
}\\
\subfloat[$\theta$-Al$_2$O$_3$/$\beta$-Ga$_2$O$_3$ interface]{ 

\includegraphics[width=8cm]{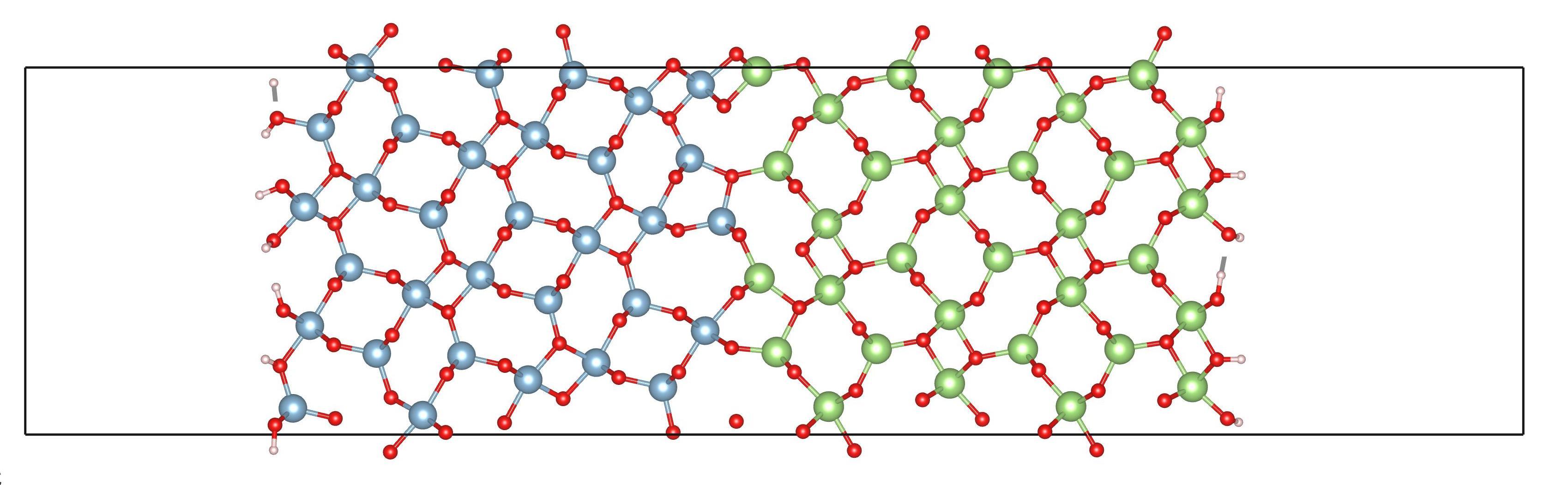} 
}
\\
\subfloat[$\gamma$-Al$_2$O$_3$/$\beta$-Ga$_2$O$_3$ interface]{
\includegraphics[width=8cm]{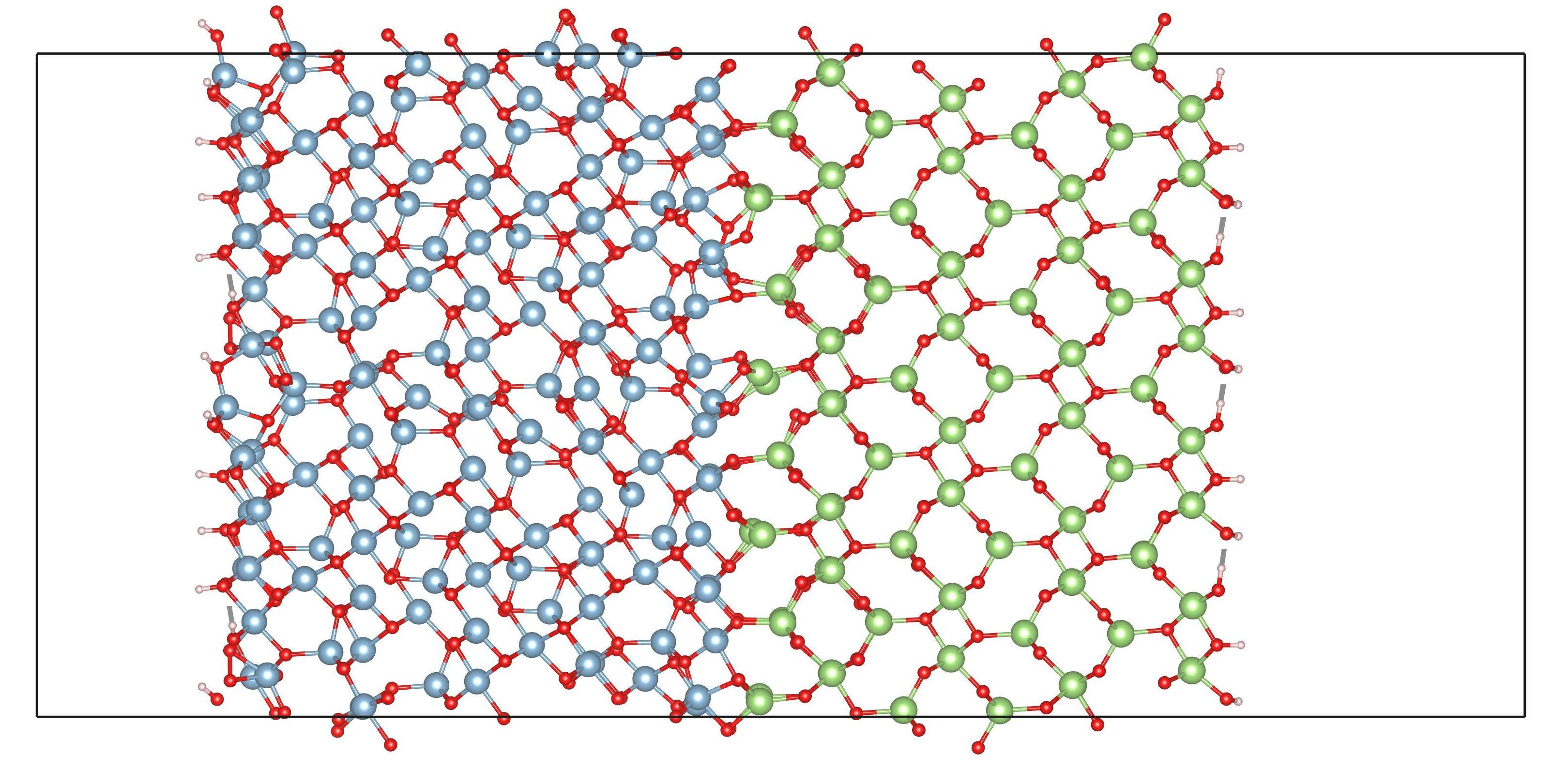} 
}
\caption{Atomistic models of the Al$_2$O$_3$/$\beta$-Ga$_2$O$_3$ interfaces obtained from structural relaxations at the GGA level. \label{intf}}
\end{figure}


In experimental studies of the band offsets between $\beta$-Ga$_2$O$_3$ and gate dielectrics, $\beta$-Ga$_2$O$_3$ is commonly taken as the substrate. Here we focus on the technologically important ($\bar{2}$01) surface of $\beta$-Ga$_2$O$_3$ for which numerous studies have been conducted to find appropriate gate dielectrics.\cite{Pearton2018} Because of the lattice mismatches between Al$_2$O$_3$ and $\beta$-Ga$_2$O$_3$,  the in-plane lattice constants of Al$_2$O$_3$ are determined by the $\beta$-Ga$_2$O$_3$ substrate. The biaxial strain due to the lattice mismatches causes the Al$_2$O$_3$ epilayer adopt new out-of-plane lattice parameters. 
To model the interface between $\alpha$-Al$_2$O$_3$ and $\beta$-Ga$_2$O$_3$, we follow the experimental determined epitaxial relationships of $\alpha$-Al$_2$O$_3$ [100] $\parallel$ $\beta$-Ga$_2$O$_3$ [102] and $\alpha$-Al$_2$O$_3$ [120] $\parallel$ $\beta$-Ga$_2$O$_3$ [010]. 
We construct an orthorhombic supercell comprising a $\alpha$-Al$_2$O$_3$ slab with (3$\times$1) in-plane periodicity and a $\beta$-Ga$_2$O$_3$ slab with (1$\times$3) in-plane periodicity. 
In the interface models, the $x$ and $y$ are parallel  to the [102] and [010] crystal axes of $\beta$-Ga$_2$O$_3$, respectively. The in-plane lattice mismatches are $-2.2\%$ and $-9.4\%$ along the $x$ and $y$ directions, respectively. The $z$ axis is perpendicular to the ($\bar{2}$01) surface of $\beta$-Ga$_2$O$_3$ for all the interface models. 
When modeling the $\kappa$-Al$_2$O$_3$/$\beta$-Ga$_2$O$_3$ interface, we use the epitaxial relationships of  $\kappa$-Al$_2$O$_3$ [100] $\parallel$ $\beta$-Ga$_2$O$_3$ [102] and $\kappa$-Al$_2$O$_3$ [010] $\parallel$ $\beta$-Ga$_2$O$_3$ [010]. The in-plane periodicities for  the $\kappa$-Al$_2$O$_3$ slab and $\beta$-Ga$_2$O$_3$ slab are ($3\times1$)  and $(1\times3)$, respectively. 
which gives rises to the lattice mismatches of $-1.9\%$ and $-8.2\%$ along the in-plane
 $x$ and $y$ directions, respectively. 
In the case of the $\theta$-Al$_2$O$_3$/$\beta$-Ga$_2$O$_3$ interface, we adopt the epitaxial relationships of $\theta$-Al$_2$O$_3$ [102] $\parallel$ $\beta$-Ga$_2$O$_3$ [102] and $\theta$-Al$_2$O$_3$ [010] $\parallel$ $\beta$-Ga$_2$O$_3$ [010]. The orthorhombic interface model contains a (1$\times$3) slab and a (1$\times$3)  $\beta$-Ga$_2$O$_3$ slab. The corresponding in-plane lattice mismatches are $-2.4\%$ and $-3.5\%$ for the $x$ and $y$ directions, respectively. 
For the $\gamma$-Al$_2$O$_3$/$\beta$-Ga$_2$O$_3$ interface, we consider the epitaxial relationships of $\gamma$-Al$_2$O$_3$ [120] $\parallel$ $\beta$-Ga$_2$O$_3$ [102] and $\gamma$-Al$_2$O$_3$ [010] $\parallel$ $\beta$-Ga$_2$O$_3$ [010]. To minimize the in-plane lattice mismatches, our orthorhombic slab is composed of a (3$\times$1) $\gamma$-Al$_2$O$_3$ slab and a (2$\times$2) $\beta$-Ga$_2$O$_3$ slab.This yields the lattice mismatches of $0.6 \%$ and $-7.4\%$ in the $x$ and $y$ directions, respectively.  For the considered four phases, the optimized lattice constants of the Al$_2$O$_3$ cells strained to the Ga$_2$O$_3$ substrate are listed in Table \ref{tabalstrain}. For $\alpha$-Al$_2$O$_3$, the strained cell is orthorhombic in which the first two lattice constants ($a$ and $b$) are same as the in-plane lattice distances in the corresponding interface model. The obtained band gap of 6.86 eV is in excellent agreement with the experimental value of 6.9 eV measured in Ref.\ \onlinecite{Carey2017}. In the case of $\theta$-Al$_2$O$_3$, the lattice constant $a$ in Table \ref{tabalstrain} denotes the in-plane distance along the [102] direction rather than along the [100] direction of the unit cell. We also provide the band gaps and the VBM positions through PBE0($\alpha$) calculations in which the mixing parameter $\alpha$ is same as that for strain-free Al$_2$O$_3$ bulk. 
\begin{figure}
\hspace*{-0.10cm}
\includegraphics[width=8cm,height=5.2cm]{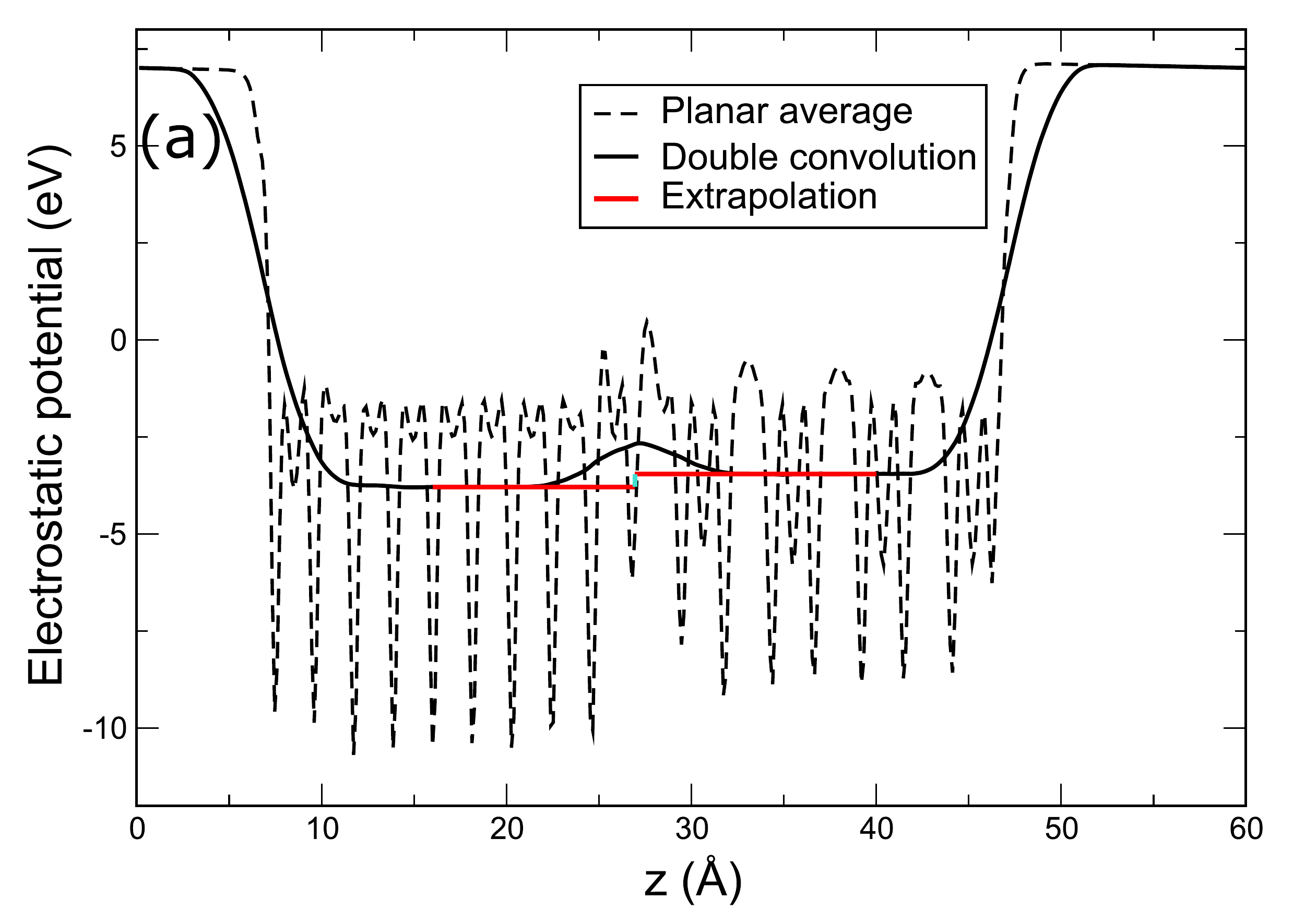}
\\ 
\hspace*{0.1cm}
\includegraphics[width=8.4cm,height=5.2cm]{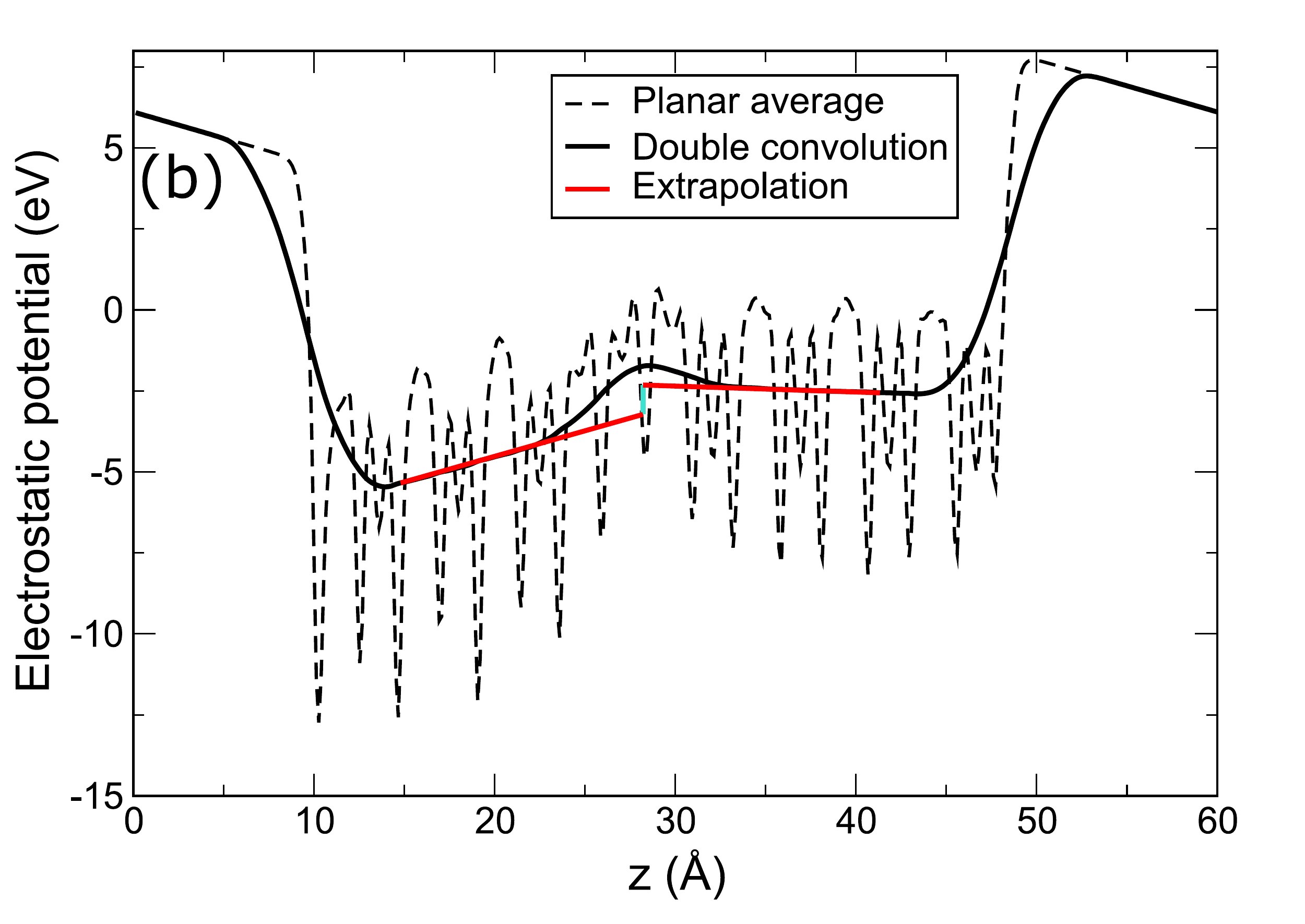}
\\
\hspace*{-0.2cm}
\includegraphics[width=8.6cm,height=5.2cm]{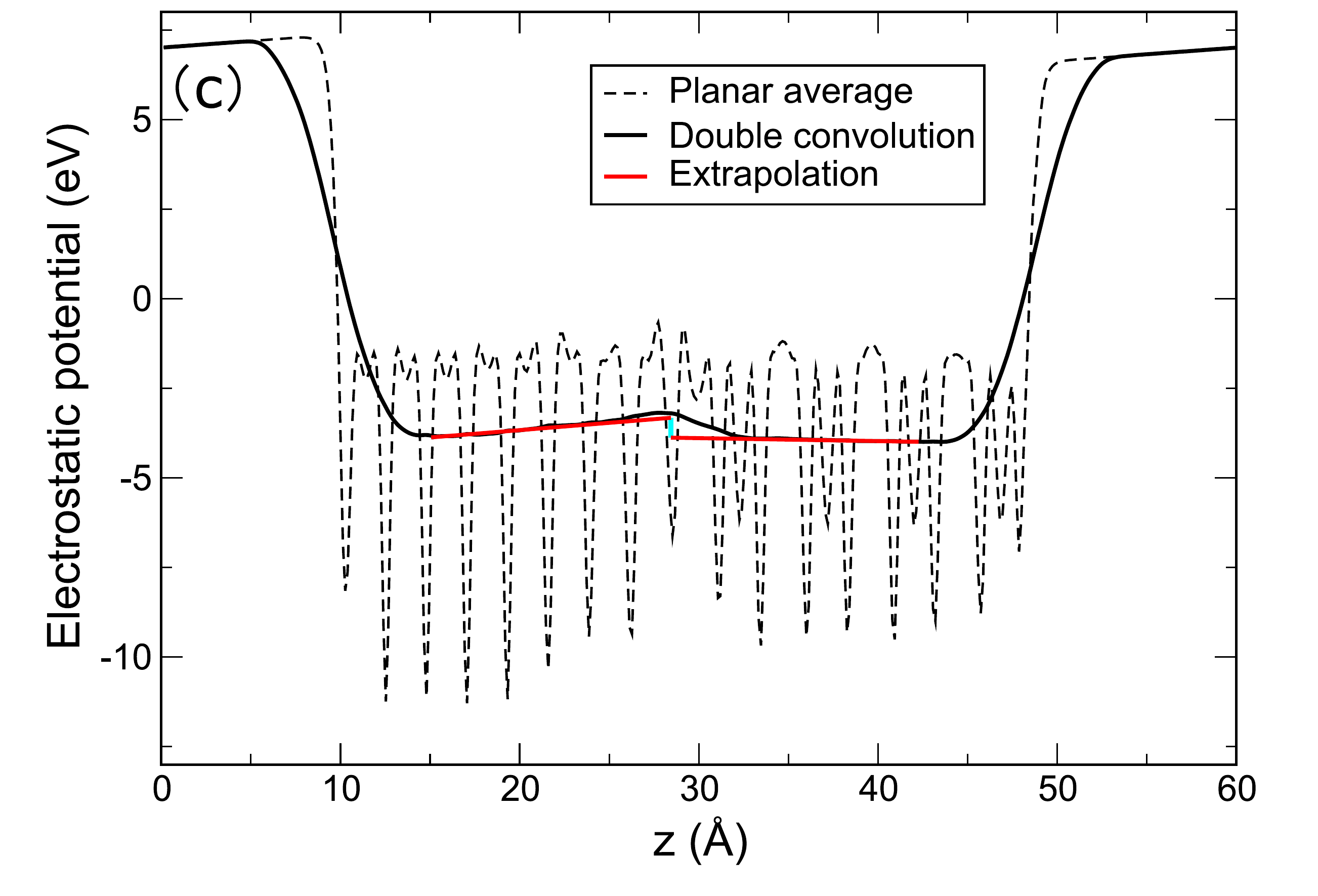}
\\
\hspace*{-0.5cm}
\includegraphics[width=8.2cm,height=5.2cm]{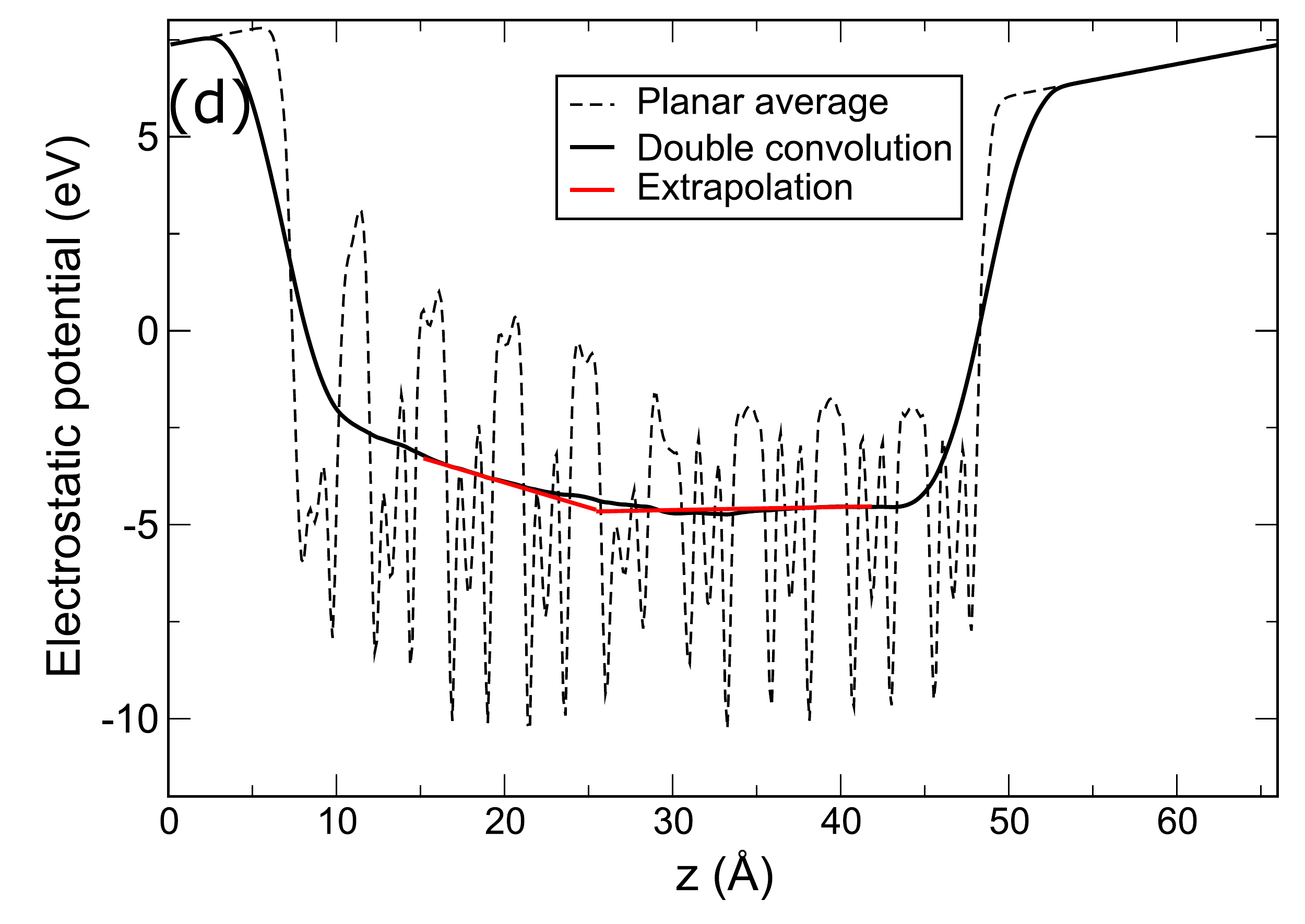}
\vskip -0.3 cm
\caption{Averaged electrostatic potential files for the (a) $\alpha$-Al$_2$O$_3$/$\beta$-Ga$_2$O$_3$, (b) $\kappa$-Al$_2$O$_3$/$\beta$-Ga$_2$O$_3$, (c) $\theta$-Al$_2$O$_3$/$\beta$-Ga$_2$O$_3$, and (d) $\gamma$-Al$_2$O$_3$/$\beta$-Ga$_2$O$_3$ interfaces calculated at the GGA level..  \label{es}}
\end{figure}

In the interface models, O atoms are used to bridge the Al$_2$O$_3$ and the $\beta$-Ga$_2$O$_3$ slabs because O atoms can allow flexibility in  bonding patterns. \cite{Lyu2020} For the surface Ga and Al atoms, there are no dangling bonds. Thick vacuum layers ($\sim$ 20 \textrm\AA) are added in the interface model ($\sim$ 60 \textrm\AA) to minimize the image interaction due to the periodic boundary conditions. Our interface models satisfy the electron-counting rule.\cite{Peacock2004,Robertson2011,Lin2011,Zhang2019}   Take the $\alpha$-Al$_2$O$_3$/$\beta$-Ga$_2$O$_3$ interface model as an example, each Ga or Al cation layer contains 12 Ga$^{3+}$ or Al$^{3+}$ ions, respectively, and each O anion layer contains 18 O$^{2+}$ ions corresponding to $-36$ charges. The surface Ga and Al layers contribute $+36$ charges and therefore exactly neutralize the interfacial O layer. The O atoms in the top and bottom layers are passiviated by the hydrogen atoms. 
After performing full structural relaxations of the atomic positions in the interface models, we calculate the electronic structures at the GGA level.
The interface models are shown in Fig.\ \ref{intf}, and
the corresponding averaged electrostatic potential profiles are shown in Fig.\ \ref{es}. We then use the alignment procedure to obtain the interface lineups.
The calculated interface lineups are given in Table \ref{taboffset}. 
\begin{table*}[ht]
\centering
\caption{Lattice parameters, band gaps (in eV) and VBM positions (in eV) of $\alpha$, $\kappa$, $\theta$, $\gamma$-Al$_2$O$_3$ and $\beta$-Ga$_2$O$_3$. The band gaps and the VBM positions are calculated at the PBE0($\alpha$) level in which the mixing parameter $\alpha$ for each material is also given.  \label{tablatt}}
\begin{ruledtabular}
\begin{tabular}{lcccccccccc}
& \mc{2}{c}{$\alpha$-Al$_2$O$_3$ }&   \mc{2}{c}{$\kappa$-Al$_2$O$_3$ }&  \mc{2}{c}{$\theta$-Al$_2$O$_3$ }&  \mc{2}{c}{$\gamma$-Al$_2$O$_3$ } &\mc{2}{c}{$\beta$-Ga$_2$O$_3$}\\ \hline
&Calc. &Exp. &Calc. &Exp. &Calc. &Exp. &Calc. &Exp. &Calc. &Exp. \\ 
 $a\ (\textrm\AA)$  & 4.80& 4.76\cite{Lucht2003}&4.88 & 4.84\cite{Halvarsson1995}& 11.88 &11.85\cite{Zhou1991}&5.67 &5.61\cite{Smrcok2006} &  12.22& 12.21\cite{Ahman1996}\\
 $b\ (\textrm\AA)$  &   &  &8.43 & 8.31& 2.95 & 2.90 & & & 3.06& 3.03\\
$c\ (\textrm\AA)$  &   13.10 &12.99 & 9.02& 8.94& 5.69&5.62 &13.94 & 13.75& 5.82&5.79\\

$\beta$ &  &  & & & 104.14$^{\circ}$ & 103.83$^{\circ}$ & & & 103.84$^{\circ}$ & $103.83^{\circ}$\\  

PBE0($\alpha$)  &0.30 & &0.29 & & 0.29 & &0.39 & &0.27\\
$E_{\textrm {gap}}^{\textrm {direct}}$ (eV) & 8.78&8.8\cite{French1990}&  7.67 & &7.61 &&7.60&7.6\cite{Filatova2015} &4.81&4.76\cite{Matsumoto1974}\\
$E_{\textrm {gap}}^{\textrm {indirect}}$ (eV) & & & & &7.22& && &4.80\\
VBM &4.08  & &3.86&&3.04& &3.20 &&2.94 \\

\end{tabular}

\end{ruledtabular}
\raggedright
\end{table*}

We also consider the situation in which Al$_2$O$_3$ is used as the substrate. 
The strain effects on the lattice constants.  and the corresponding band gaps, and the VBM levels of $\beta$-Ga$_2$O$_3$ have to be accounted for. To achieve this, 
We convert the conventional unit cell of $\beta$-Ga$_2$O$_3$ into a larger monoclinic one with the ($\bar{2}$01) face.  The mathematical relationship between the first lattice vector \textbf{{a}$'$} of the larger cell and \textbf{a} of the conventional unit cell can be represented by the equation: \textbf{{a}$'$}=\textbf{a}+2\textbf{{c}}. The other two lattice vectors remain unchanged and the angle between  \textbf{{a}$'$}  and \textbf{c} is denoted as $\beta'$. 
This larger unit cell is strained to the Al$_2$O$_3$ substrate which determines the in-plane lattice constants ($a'$ and $b$). The other lattice parameters ($c$ and $\beta'$) and the internal coordinates are optimized through structural relaxations.
The calculated lattice parameters,
band gaps, and the VBM levels of $\beta$-Ga$_2$O$_3$ subject to different Al$_2$O$_3$ substrates are given in Table \ref{tabgastrain}. The exchange mixing parameter of $\alpha=0.27$ is the same as that for the unstrained bulk. 
The band gaps of the strained $\beta$-Ga$_2$O$_3$ cells at compressed volumes are larger than that of the unstrained bulk, which is consistent with the deformation potentials of $\beta$-Ga$_2$O$_3$. \cite{Swallow2021} For the $\beta$-Ga$_2$O$_3$/Al$_2$O$_3$ interfaces, we perform structural relaxations and then determine the corresponding interface lineups as summarized in Table \ref{taboffset}.

\begin{table}
\centering
\caption{Lattice parameters, band gaps and VBM positions relative to the bulk reference levels of the strained Al$_2$O$_3$ cells on the $\beta$-Ga$_2$O$_3$ substrate. \label{tabalstrain}}
\begin{ruledtabular}
\begin{tabular}{lcccc}

Strained& \mc{1}{c}{$\alpha$-Al$_2$O$_3$ }&   \mc{1}{c}{$\kappa$-Al$_2$O$_3$ }&  \mc{1}{c}{$\theta$-Al$_2$O$_3$ }&  \mc{1}{c}{$\gamma$-Al$_2$O$_3$ }\\ 
Substrate &  $\beta$-Ga$_2$O$_3$   &$\beta$-Ga$_2$O$_3$    & $\beta$-Ga$_2$O$_3$  & $\beta$-Ga$_2$O$_3$  \\  \hline
 $a\ (\textrm\AA)$  & 4.91 & 4.91 &14.72$^a$ & 9.81\\
 $b\ (\textrm\AA)$  &  9.18  & 9.18 & 3.06 & 12.24\\

$c\ (\textrm\AA)$  &   12.81 & 8.80 & 5.71 & 13.73\\
$\beta$ &  &  & $128.17^{\circ}$\\

PBE0($\alpha$) & 0.30 & 0.29 & 0.29 &0.39 \\
$E_{\textrm {gap}}^{\textrm {direct}}$ (eV) &6.86 & 6.67 & 6.94 & 6.89\\
$E_{\textrm {gap}}^{\textrm {indirect}}$ (eV) &    &        &   6.90       & \\
VBM & 3.57 & 3.35 & 2.97 & 2.56 \\

\end{tabular}
\end{ruledtabular}
$^a$ This is the lattice constants along the [102] direction. 
\end{table}

\begin{table}
\centering
\caption{Lattice parameters, band gaps, and VBO levels with respect to the bulk reference levels of the strained $\beta$-Ga$_2$O$_3$ on the Al$_2$O$_3$ substrates. \label{tabgastrain}}
\begin{ruledtabular}
\begin{tabular}{lcccc}
Strained &$\beta$-Ga$_2$O$_3$   &$\beta$-Ga$_2$O$_3$    & $\beta$-Ga$_2$O$_3$  & $\beta$-Ga$_2$O$_3$  \\ 
Substrate & \mc{1}{c}{$\alpha$-Al$_2$O$_3$ }&   \mc{1}{c}{$\kappa$-Al$_2$O$_3$ }&  \mc{1}{c}{$\theta$-Al$_2$O$_3$ }&  \mc{1}{c}{$\gamma$-Al$_2$O$_3$ }\\ \hline

 $a'\ (\textrm\AA)$  & 14.41 & 14.64 & 14.37 &14.80\\
$b\ (\textrm\AA)$  &  2.79 & 2.81 & 2.95 &2.83\\
$c\ (\textrm \AA)$ & 5.83 & 5.81 & 5.83 & 5.80  \\

$\beta'$ & 56.07$^{\circ}$ &55.73$^{\circ} $ & 54.97$^{\circ}$ &55.33$^{\circ}$ \\

PBE0($\alpha$)  & 0.27 & 0.27 & 0.27 & 0.27\\
$E_{\textrm {gap}}^{\textrm {direct}}$ (eV) &5.35 &5.32 & 5.15 & 5.28 \\
VBM &3.88  & 3.63& 3.42 & 3.46\\

\end{tabular}

\end{ruledtabular}
\end{table}

\begin{table}
\centering
\caption{Calculated inteface lineup (in eV) and band offsets (in eV) at the interfaces between $\beta$-Ga$_2$O$_3$ and Al$_2$O$_3$. The available experimental and theoretical results are also given.  \label{taboffset}}
\begin{ruledtabular}
\begin{tabular}{lccc}
Interface & Interface lineup & VBO &CBO \\ \hline
$\alpha$-Al$_2$O$_3$/$\beta$-Ga$_2$O$_3$ &   -0.33 &0.30 &2.36   \\
Expt.  \cite{Kamimura2014}                                                           &  & -0.70 & 1.50 \\
Expt. \cite{Carey2017}                                                               &  &-0.07  & 2.23 \\
Expt.    \cite{Carey2017}                          & & 0.86 & 3.16 \\
Expt. \cite{Hung2014}                              &  &  &1.7 \\ 
Calc. \cite{Peelaers2018}              &  & -0.27 & 3.68 \\
$\kappa$-Al$_2$O$_3$/$\beta$-Ga$_2$O$_3$  &  -0.91  & -0.40& 1.46\\
$\theta$-Al$_2$O$_3$/$\beta$-Ga$_2$O$_3$  &0.55 &   0.54 &  2.63\\
Calc. \cite{Peelaers2018}                                &   & 0.37 &2.74 \\
$\gamma$-Al$_2$O$_3$/$\beta$-Ga$_2$O$_3$  & 0.04& -0.34 & 1.74    \\
Expt. \cite{Hattori2016}   & &  -0.5 & 1.9 \\
\hline
$\beta$-Ga$_2$O$_3$/$\alpha$-Al$_2$O$_3$ &   1.17 & 0.97&-2.46 \\
$\beta$-Ga$_2$O$_3$/$\kappa$-Al$_2$O$_3$ &  0.19 & -0.04 &-2.39\\
$\beta$-Ga$_2$O$_3$/$\theta$-Al$_2$O$_3$ & -0.24&   0.14&  -2.32\\
$\beta$-Ga$_2$O$_3$/$\gamma$-Al$_2$O$_3$ & -0.23   & 0.03  & -2.29\\

\end{tabular}
\end{ruledtabular}
\end{table}

\begin{figure}
\hspace*{-0.60cm}
\subfloat[Al$_2$O$_3$/$\beta$-Ga$_2$O$_3$ interfaces]{
\includegraphics[width=9.6cm]{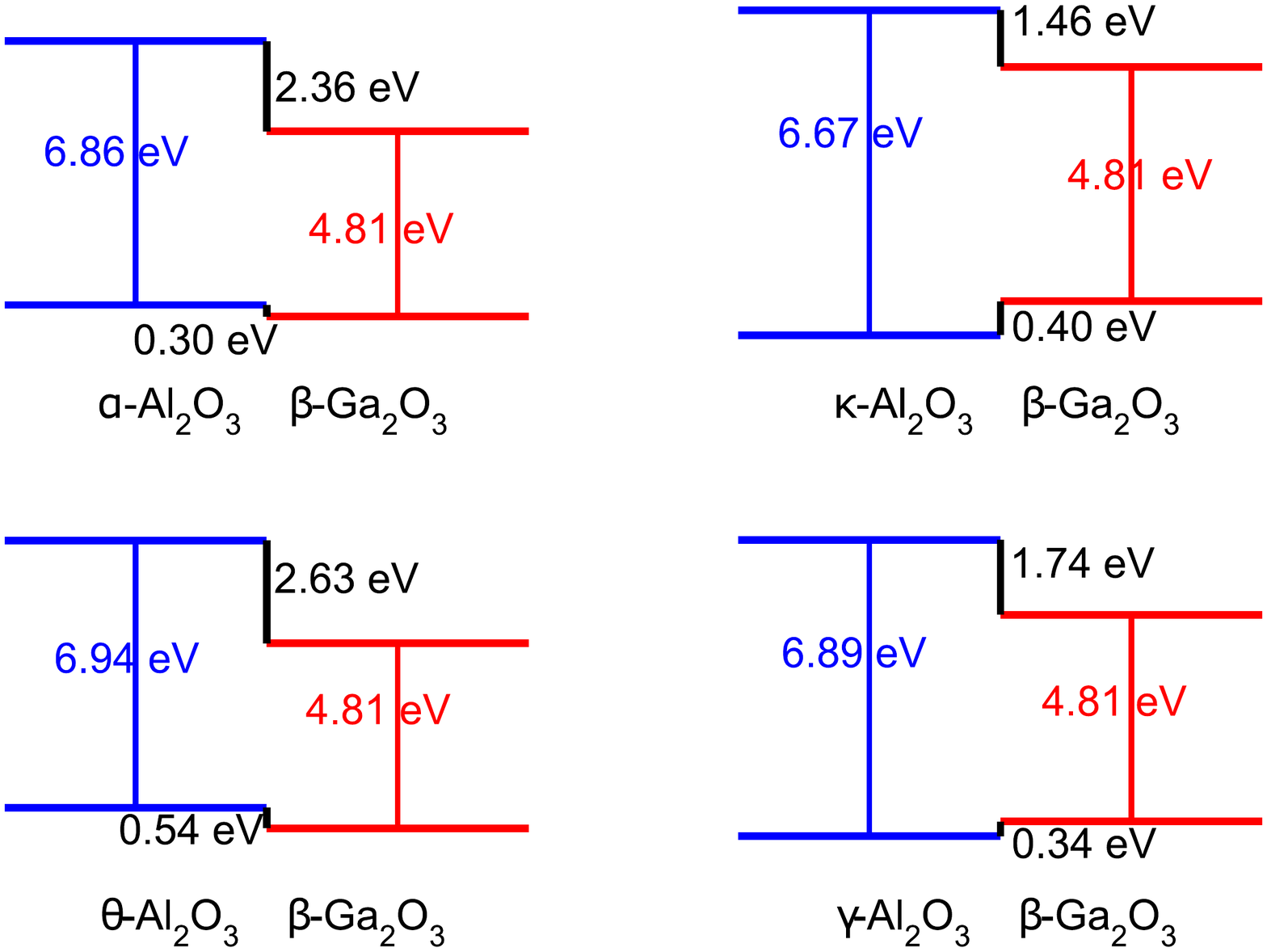}
}
\\
\vskip -0.2cm
\hspace*{-0.60cm}
\subfloat[$\beta$-Ga$_2$O$_3$/Al$_2$O$_3$ interfaces]{
\includegraphics[width=9.6cm]{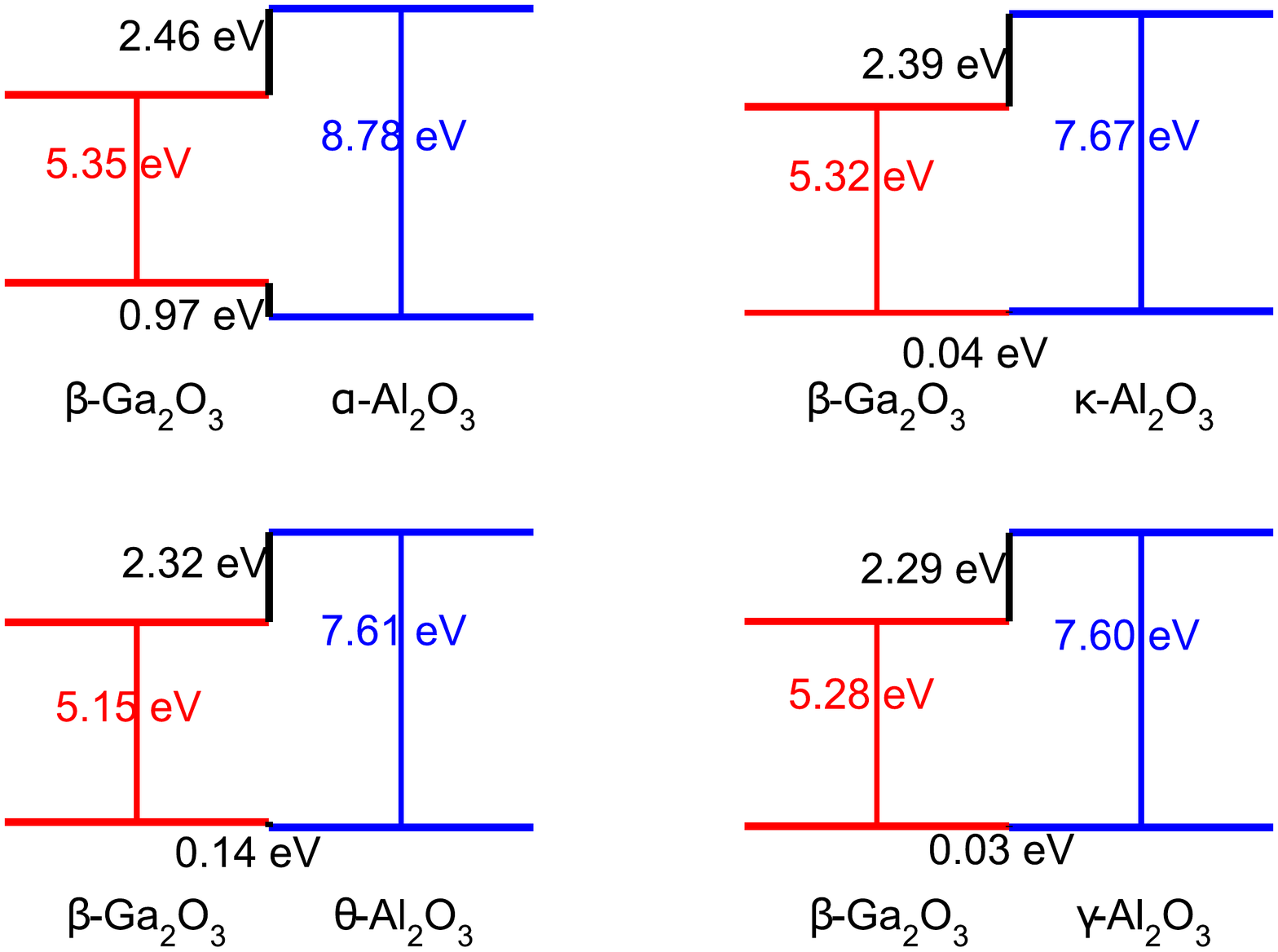} 
}
  \vskip -0 cm
\caption{Band alignment diagrams of the interfaces between $\beta$-Ga$_2$O$_3$ and Al$_2$O$_3$. In (a) and (b), the substrates are Al$_2$O$_3$ and Al$_2$O$_3$, respectively. The signs of the offsets are dropped out for brevity. 
\label{figoffset}}
\end{figure}

The calculated band offsets  together with the available experimental and theoretical results at the interfaces between $\beta$-Ga$_2$O$_3$ and Al$_2$O$_3$ are given in Table \ref{taboffset}. Note the signs of the literature results are adjusted according to the definitions of VBO and CBO in the Methods section. In Fig.\ \ref{figoffset}, we schematically show the calculated valence and conduction band offsets. For $\alpha$-Al$_2$O$_3$ on $\beta$-Ga$_2$O$_3$, the calculated VBO of 0.7 eV and CBO of 2.36 eV favor the middle of the range of the experimentally measured offsets.\cite{Carey2017, Kamimura2014, Hung2014}  Recently, 
Peelaers \etal calculated a VBO of $-$0.27 eV and a CBO of 3.68 eV between unstrained $\alpha$-Al$_2$O$_3$ and $\beta$-Ga$_2$O$_3$ bulks from the electron affinity rule.\cite{Peelaers2018} The differences from our results partly because that we explicitly consider the strain effects. 
For $\kappa$-Al$_2$O$_3$/$\beta$-Ga$_2$O$_3$, we obtain a VBO of $-$0.40 eV and a CBO of 1.46 eV. 
In the case of the $\theta$-Al$_2$O$_3$/interface, the calculated VBO and CBO are 0.54 eV and 2.63 eV, respectively. 
For this interface,  
Peelaers \etal calculated a VBO of 0.37 eV and a CBO of 2.74 eV between unstrained $\theta$-Al$_2$O$_3$ and $\beta$-Ga$_2$O$_3$ by assuming the electron affinity rule. \cite{Peelaers2018} We suggest this good agreement arises from the fact that the lattice mismatches between $\theta$-Al$_2$O$_3$ and $\beta$-Ga$_2$O$_3$ are rather small and the interface model satisfies the electron counting rule. 
For $\gamma$-Al$_2$O$_3$ on  $\beta$-Ga$_2$O$_3$, the calculated VBO of $-$0.34 eV is in good agreement with the experimental value of $-$0.5 eV for $\gamma$-Al$_2$O$_3$ on (010) $\beta$-Ga$_2$O$_3$ reported by Hattori \etal\cite{Hattori2016}.  We suggest this  agreement is partly because of  the satisfaction of the electron counting rule in our models despite the fact that the surfaces involved of $\beta$-Ga$_2$O$_3$ are different.  For Al$_2$O$_3$ on $\beta$-Ga$_2$O$_3$, the $\alpha$ and $\theta$ phases form type II heterojunctions.  For $\kappa$-Al$_2$O$_3$ and $\gamma$-Al$_2$O$_3$ on ($\bar{2}01$)$\beta$-Ga$_2$O$_3$, there are type I band alignments but the corresponding VBOs are less than 1 eV, thereby indicating not very sufficient barriers for holes. 

We then discuss the band alignments of $\beta$-Ga$_2$O$_3$ on Al$_2$O$_3$. For $\beta$-Ga$_2$O$_3$/$\alpha$-Al$_2$O$_3$, we obtain a VBO of 0.97 eV and a CBO of $-$2.46 eV. Both the CBO and the VBO are $\gtrsim$ 1 eV, therefore we identify $\alpha$-Al$_2$O$_3$ as an appropriate candidate for gate dielectrics on $\beta$-Ga$_2$O$_3$ in MOSFETs. For $\beta$-Ga$_2$O$_3$ on $\kappa$-, $\theta$-, and $\gamma$-Al$_2$O$_3$, we find that the VBOs are nearly negligible but the CBOs are $\sim$ 2.4 eV indicating sufficient barriers for electrons. Hence, these three phase of Al$_2$O$_3$ can be used as electron blocking layers in $\beta$-Ga$_2$O$_3$-based LEDs. \cite{Lin2020}

In conclusion, we studied the band offsets at the interfaces between ($\bar{2}$01) $\beta$-Ga$_2$O$_3$ and the four representative phases of Al$_2$O$_3$ ($\alpha$, $\kappa$, $\theta$, and $\gamma$) through the state-of-the-art hybrid density functional calculations. The calculated band offsets are in line with the available experimental results.
The modeling procedures in this study can directly be applied to the interfaces between $\beta$-Ga$_2$O$_3$ and technologically attractive monoclinic (Al$_x$Ga$_{1– x}$)$_2$O$_3$ alloys\cite{Peelaers2018, Amol2020}, and be useful for the study of the interfaces involving (010) $\beta$-Ga$_2$O$_3$. 
More generally, the present study shows how to address band alignments at the interfaces between $\beta$-Ga$_2$O$_3$ with oxides. The calculated band alignments are essential for the device designs based on $\beta$-Ga$_2$O$_3$ such as MOSFETs and LEDs.

\section*{Acknlowledgements}
Sai Lyu would like to thank Prof. Walter R. L. Lambrecht for the fruitful discussions. The author acknowledges financial support from Shandong Normal University. This work is supported by National Supercomputer Center in Guangzhou.

\section*{DATA AVAILABILITY}
The data that support the findings of this study are available
from the corresponding author upon reasonable request.

\bibliography{ga2o3}

\end{document}